# Static magnetic order with strong quantum fluctuations in spin-1/2 honeycomb magnet Na$_2$Co$_2$TeO$_6$

Gaoting Lin[1#], Jinlong Jiao[1#], Xiyang Li[2#], Mingfang Shu[1], Oksana Zaharko[3], Toni Shiroka[4,5], Tao Hong[6], Alexander I. Kolesnikov[6], Guochu Deng[7], Sarah Dunsiger[8,9], Haidong Zhou[10], Tian Shang[11*], and Jie Ma[1*]

[1]*Key Laboratory of Artificial Structures and Quantum Control, Shenyang National Laboratory for Materials Science, School of Physics and Astronomy, Shanghai Jiao Tong University, Shanghai 200240, China*

[2]*Quantum Matter Institute, University of British Columbia, Vancouver, British Columbia V6T 1Z4, Canada*

[3]*Laboratory for Neutron Scattering and Imaging, Paul Scherrer Institut, Villigen, Switzerland*

[4]*Laboratory for Muon-Spin Spectroscopy, Paul Scherrer Institut, Villigen PSI, Switzerland*

[5]*Laboratorium für Festkörperphysik, ETH Zürich, CH-8093 Zürich, Switzerland*

[6]*Neutron Scattering Division, Oak Ridge NationalLaboratory, Oak Ridge, TN 37831, USA*

[7]*Australian Centre for Neutron Scattering, Australian Nuclear Science and STechnology Organization, New Illawarra Road, Lucas Heights, NSW 2234, Australia*

[8]*Department of Physics, Simon Fraser University, Burnaby, BC, V5A 1S6, Canada*

[9]*Centre for Molecular and Materials Science, TRIUMF, Vancouver, BC, V6T 2A3, Canada*

[10]*Department of Physics and Astronomy, University of Tennessee, Knoxville, Tennessee 37996, USA*

[11]*Key Laboratory ofPolar Materials and Devices (MOE), School of Physics and Electronic Science, East China Normal University, Shanghai 200241, China*

\# These authors contributed equally to this work.
*Corresponding author:tshang@phy.ecnu.edu.cn
jma3@sjtu.edu.cn

**Abstract**

Kitaev interactions, arising from the interplay of frustration and bond anisotropy, can lead to strong quantum fluctuations and, in an ideal case, to a quantum-spin-liquid state. However, in many nonideal materials, spurious non-Kitaev interactions typically promote a zigzag antiferromagnetic order in the $d$-orbital transition metal compounds. By combining neutron scattering with muon-spin rotation and relaxation techniques, we provide new insights into the exotic properties of $Na_2Co_2TeO_6$, a candidate Kitaev material. Below $T_N$, the zero-field muon-spin relaxation rate becomes almost constant (at 0.45 μs$^{-1}$). We attribute this temperature-independent muon-spin relaxation rate to the strong quantum fluctuations, as well as to the frustrated Kitaev interactions. As the magnetic field increases, neutron scattering data indicate a much broader spin-wave-excitation gap at the $K$-point. Therefore, quantum fluctuations seem not only robust, but are even enhanced by the applied magnetic field. Our findings provide valuable hints for understanding the onset of the quantum-spin-liquid state in Kitaev materials.

The Kitaev model, introduced in a spin-1/2 two-dimensional (2D) honeycomb lattice as an exactly solvable 2D spin model that achieves a quantum-spin-liquid (QSL) ground state [1], has attracted considerable attention. Numerous theoretical and experimental attempts have been made in the search for a QSL candidate with dominant bond-dependent anisotropic-exchange interactions $K$, named Kitaev interactions [2-18]. To date, the Kitaev model has been successfully realized in several 3$d$, 4$d$, and 5$d$ transition metal families [8-13,18-21]. Unfortunately, owing to the ubiquitous presence of non-Kitaev interactions [6,7,13,15], e.g., Heisenberg exchanges, or off-diagonal symmetric interactions $\Gamma$ and $\Gamma'$, at low temperatures and in the absence of magnetic fields, most of these materials fail to realize a QSL state and exhibit instead a zigzag antiferromagnetic (AFM) order [6-13,18-21].

Compared to the previously discovered 4$d$/5$d$ Ru/Ir systems, characterized by a strong spin-orbit coupling (SOC) [15-17], the 3$d$ Co-based Kitaev QSL candidate materials remain highly controversial [12,13,19,22-28]. A good example of such materials is $Na_2Co_2TeO_6$ (NCTO), originally thought as one of the most prominent cases for studying Kitaev physics [12,13,15,19,28-35]. Although the evidence of a field-induced QSL is available [13,19,36,37], the magnetic structure of its ground state remains an open and intriguing question [13,19,26,27,31,33,38], whose answers should help us understand the role of the competitors of $K$, such as the nearest-neighbor (NN) Heisenberg coupling $J_1$ and the third NN Heisenberg coupling $J_3$.

Recently, by using various experimental methods, such as single-crystal or powder neutron diffraction, nuclear magnetic resonance, and electrical polarization measurements, researchers have attempted to understand the magnetic structure of NCTO [26,27,29-31,39,40], yet an undisputed conclusion has not been reached. i) Powder neutron diffraction experiments suggest a zigzag AFM order, with the magnetic moments lying in the $ab$ plane [see Fig. 1(a)], accompanied by a Néel-type canting along the $c$ axis below $T_N \sim 26$ K [39,40]. However, single-crystal neutron diffractions suggest a triple-$q$ order (or multi-$k$ structure) in the absence of a magnetic field [12,27,28,31]. ii) Although the field dependence of the characteristic magnetic reflections (0.5, 0, 1) and (0, 0.5, 1) demonstrate the magnetic multi-domain effect of the zigzag AFM order [27], the temperature dependence of the (0.5, 0, 0) magnetic reflection cannot rule out a triple-$q$ order [12,27,31]. iii) The spin dynamics was analyzed by a generalized Heisenberg–Kitaev model with five symmetry-allowed terms: $K$, $\Gamma$ and $\Gamma'$, $J_1$, and $J_3$ [13,33-35,41], and two different exchange frustrations: Kitaev and $J_1$-$J_3$. However, such a complex model made the determination of the ground state rather difficult. To distinguish the two possible magnetic structures, i.e., multi-domain or multi-$k$, a good strategy consists of measuring the field- and temperature-dependent magnetic reflections and the related dynamics at the $M$-points lying in the $ab$-plane, such as M (0.5, 0, $L$), $M_1$ (-0.5, 0.5, $L$), and $M_2$(0, 0.5, $L$), where $L$ is an arbitrary integer along the $c$-axis [see Fig. 1(b)].

In this letter, we investigate the zigzag AFM order of NCTO single crystals via neutron scattering and muon-spin rotation and relaxation (μSR) techniques. The change of the characteristic magnetic reflections $M$ (0.5, 0, 1) and $M_1$ (-0.5, 0.5, 1) with temperature and magnetic field indicate that the magnetic ground state presents a multi-domain structure, rather than a multi-$k$ structure. Below $T_N$, the modulation of the magnetic domains induces other two transitions at $T_F$ and $T^*$, reflected also in the temperature-dependent weak transverse-field (wTF) μSR asymmetry and the zero-field (ZF) muon-spin relaxation rate. Furthermore, the estimated static magnetic volume fraction (~ 90%) confirms the high quality of the NCTO single crystals. The relatively large and temperature-independent muon-spin relaxation rate in the AFM state suggests the presence of strong quantum spin fluctuations in NCTO.

The measurements were performed on high-quality single crystals grown by the flux method, as described elsewhere in Ref. [19]. A piece of NCTO single crystal was used for the neutron diffraction experiments [see the inset in Fig. S1(a)], performed at the thermal single-crystal diffractometer ZEBRA at the Swiss Spallation Neutron Source SINQ, Paul Scherrer Institut (PSI), Switzerland. A neutron wavelength of 1.383 Å, obtained by a Ge monochromator, was used for all the measurements at ZEBRA. In-field data were collected using a lifting arm normal-beam geometry, where the crystal was inserted in a 10-T vertical magnet [42]. The magnetic states of NCTO were investigated with ***B*** // [-1 1 0] (equivalent to ***a*****-axis). Spin-wave excitation spectra were measured using the SEQUOIA time-of-flight spectrometer at the Spallation Neutron Source, Oak Ridge National Laboratory, USA [43,44]. The constant-*Q* scans of the *K*-point (1/3, 1/3, 0) were measured at 1.5 K with magnetic fields up to 4 T on the Cold Triple Axis Spectrometer SIKA at the ANSTO, Australia [45]. The ZF-, longitudinal-field (LF-), and wTF-μSR experiments were performed on the M20D surface muon beam line at TRIUMF in Vancouver, Canada using the LAMPF spectrometer [42]. In addition, wTF-μSR experiments were also carried out at the general-purpose surface-muon (GPS) instrument at the πM3 beamline of the Swiss muon source (SμS) at PSI in Villigen, Switzerland [42]. All the μSR spectra were analyzed using the musrfit software package [46].

To determine whether the magnetic ground state of NCTO is a multi-domain structure or a multi-*k* structure, both the temperature- and magnetic-field dependence of the magnetic reflections *M* (0.5, 0, 1) and $M_1$ (-0.5, 0.5, 1) were measured for the wave vectors *k* = (0.5, 0, 0) and (-0.5, 0.5, 0), respectively [see Fig. 1(c) and (d)]. The differences in the non-normalized intensity magnitudes at the *M* and $M_1$ points may be due to the non-spherical sample morphology (see sample photo in Fig. S1 [42]). The temperature-dependent intensity curves *I*(*T*) at both *M* and $M_1$ points can be well fitted using a phenomenological model: $I(T) = I_0 \cdot [1-(T/T_N)^2]^\beta$ [47], yielding a magnetic ordering temperature $T_N \approx 26$ K and $\beta \approx 0.22$. The *I*(*T*) curves for both *M* and $M_1$ show similar behaviors, but a small bump near $T_F \approx 15$ K only appears for the $M_1$ point, implying that the anomaly at $T_F$ may not be attributed to the magnetic phase transition. Interestingly, the spin-wave-excitation gap decreases with increasing temperature and vanishes exactly at temperatures close to $T_F$ [31]. The appearance of the spin-wave-excitation gap can be attributed to the SOC inducing magnetic anisotropic-exchange interactions *K* and *Γ* terms in NCTO [13], which may lead to the non-equivalent magnetic domains with different *k* vectors at the *M* and $M_1$ points. Hence, the subtle differences between these two magnetic reflections indicate that the magnetic ground state is a multi-domain structure, that is, the domains corresponding to the *k* vectors of (0.5, 0, 0) and (-0.5, 0.5, 0) are not equally populated. In addition, the estimated $\beta \approx$ 0.22 is neither consistent with the ideal 2D Ising ($\beta = 0.125$) nor the 3D Ising ($\beta = 0.326$) system [48], but it is more consistent with the quasi-2D magnetic correlation [13,19,31].

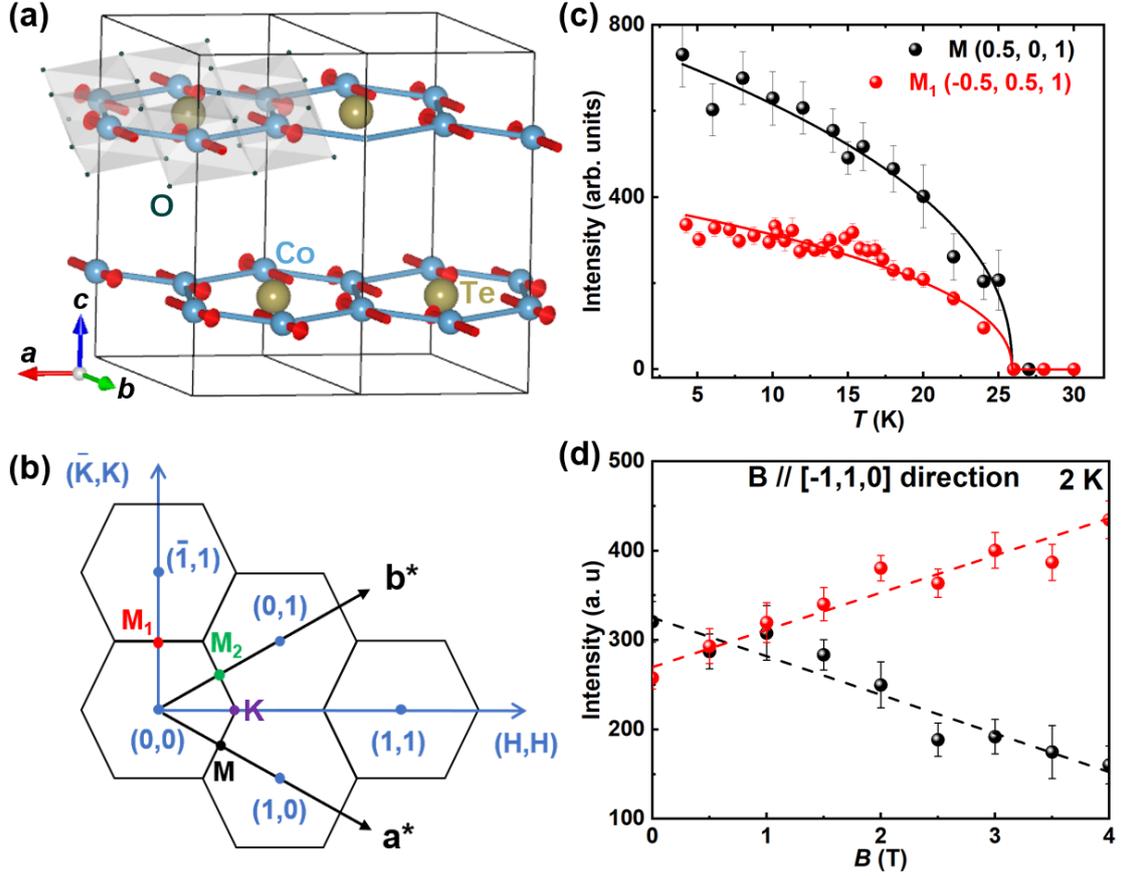

Figure 1. **The honeycomb structure and single-crystal neutron diffraction for NCTO.** (a) Magnetic structure of NCTO. (b) Schematic plots of the Brillouin zones showing the in-plane high symmetry points $M$, $M_1$, $M_2$, and $K$ denoted by black, red, green, and purple dots, respectively. The high symmetry points are identical for the $L$ with different integers along the $c$-axis. (c) The temperature dependence of the zero-field intensity of two magnetic reflections at $M$ (0.5, 0, 1) and $M_1$ (-0.5, 0.5, 1). The solid lines represent the fits to the phenomenological model (see details in the text). (d) Field-dependent intensity of magnetic reflections at $M$ and $M_1$ points, collected at $T = 2$ K and with the magnetic field applied along [-1, 1, 0] direction. Dashed lines are guides to the eyes.

As shown in Fig. 1(d), the multi-domain structure is further supported by the field-dependent intensity $I(B)$ at the magnetic reflections $M$ and $M_1$ for ***B*** // [-1, 1, 0]. The $M$ and $M_1$ reflections exhibit completely opposite field dependence. When increasing the magnetic field, the intensity of the *M* reflection decreases, while that of $M_1$ increases. Such opposite field dependence implies that the macroscopic symmetry is broken by the applied magnetic field and, thus, that the magnetic domains along the field direction gradually grow, while the domains in other directions are suppressed [49]. With the application of an external constraint (magnetic field or uniaxial stress), each arm of the *k* vector would exhibit a different response in case of a multi-domain scenario, while they would show a similar behavior in case of a multi-*k* scheme [50-53]. Clearly, our neutron data support a multi-domain structure rather than a multi-*k* structure in NCTO single crystals. It is worth mentioning that, in the zigzag AFM state, NCTO exhibits a twofold symmetry in the angular dependence of the magnetic torque [19], which might be related to the different *I(B)* field responses at the magnetic reflections *M* and $M_1$.

Another feature of quantum fluctuations in certain strong quantum magnets is a broadening of the spin-wave-excitation spectra [54-56]. To search for the quantum fluctuations in an external magnetic field, we performed single-crystal inelastic-neutron-scattering measurements by applying various magnetic fields up to 4 T along the [-1,1,0] direction. As shown in Figs. 2(a) and (b), a spin-wave band can be identified at 0 and 4 T along the high symmetry momentum directions *Γ-K-M* (see the arrow in the inset of Fig. 2). As expected, in the zero-field case, the gapped magnon band at the *M*-point reaches the minimum energy value and has the largest intensity, thus supporting a zigzag AFM order driven by the non-Kitaev interactions [12,19]. The spin-wave band at 4 T shows similar features to the lowest-energy spin-wave-excitation spectra at 0 T [see Fig. 2(b)], but it is significantly broadened by the external field. Note that, in NCTO, we do not observe a splitting of the spin-wave band into multiple branches under an applied magnetic field. Hence, such broadening is most likely caused by the field-induced quantum fluctuations, which can be further evidenced by constant-*Q* scans at the *K*-point. As shown in Fig. 2(c), at 0 T, the larger excitation gap is centered around 3 meV [12,19]. Upon increasing the magnetic field to 4 T, the zigzag AFM order is still present [see *I(B)* curves in Fig. 1(d)], but the excitation gap broadens continuously. The full width at half maximum (FWHM) of the excitation gap vs the magnetic field is shown in the inset of Fig. 2(c). The above results indicate that quantum fluctuations in NCTO are enhanced by an applied magnetic field. This is highly consistent with the field-induced magnetically disordered state with strong quantum fluctuations observed between 7.5 and 10 T with ***B*** // ***a\****-axis [13,19,37].

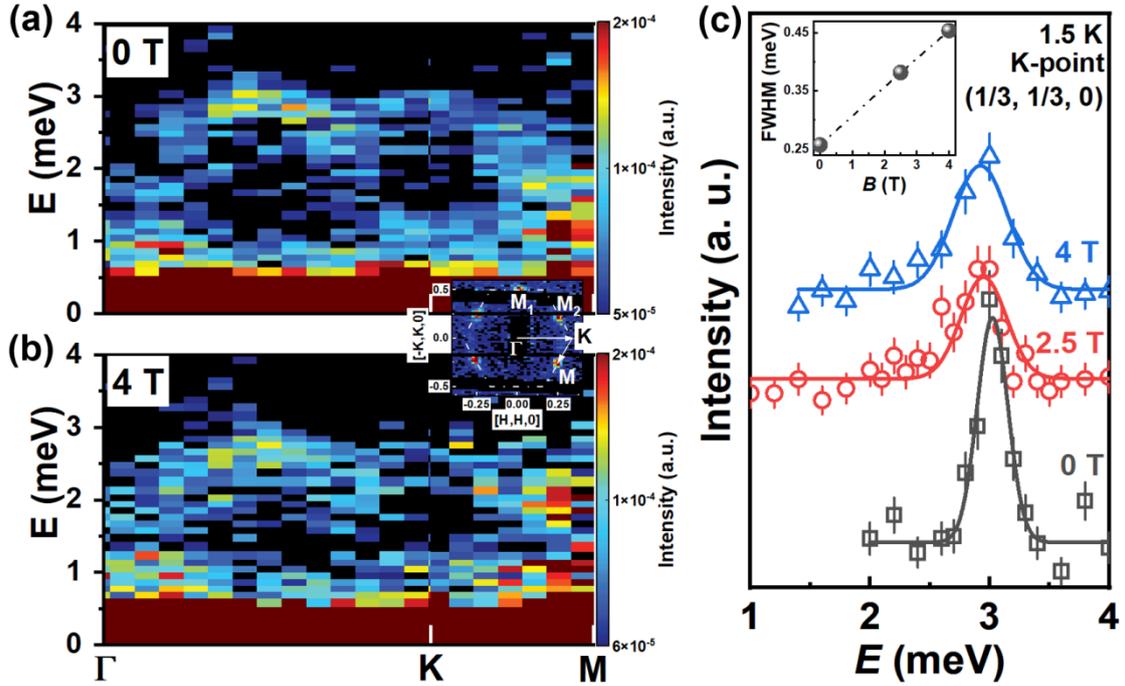

Figure 2. **Single-crystal inelastic neutron scattering results for NCTO**. (a) and (b) Lowest-energy spin-wave-excitation spectra at 2 K for 0 T and 4 T, respectively. The color bars indicate the scattering intensity in a linear scale. The inset shows the elastic neutron scattering integrated at the elastic location $L = [-2.5, 2.5]$ and $E = [-0.075, 0.075]$ meV at 0 T using fixed incident energy $E_i = 60$ meV. The white dash-dotted lines represent the Brillouin zone boundaries. The high symmetry points $\Gamma$, $K$, $M$, $M_1$ and $M_2$ are marked and the white arrows show the high symmetry momentum directions $\Gamma$-$K$-$M$ path in the inset. (c) The constant-$Q$ scans of the $K$-point (1/3, 1/3, 0) were collected at $T = 1.5$ K by applying magnetic fields up to 4 T along the [-1,1,0] direction. The solid lines represent the fits using a Gaussian function. The inset shows the field dependence of the FWHM of the excitation gap.

As an extremely sensitive probe of complex quantum magnetism, μSR is regularly used to investigate the magnetic order at a microscopic level. Here, the combination of a long-range (neutron scattering) and a short-range (μSR) technique helped us to confirm the multi-domain structure and the quantum fluctuations in a NCTO single crystal.

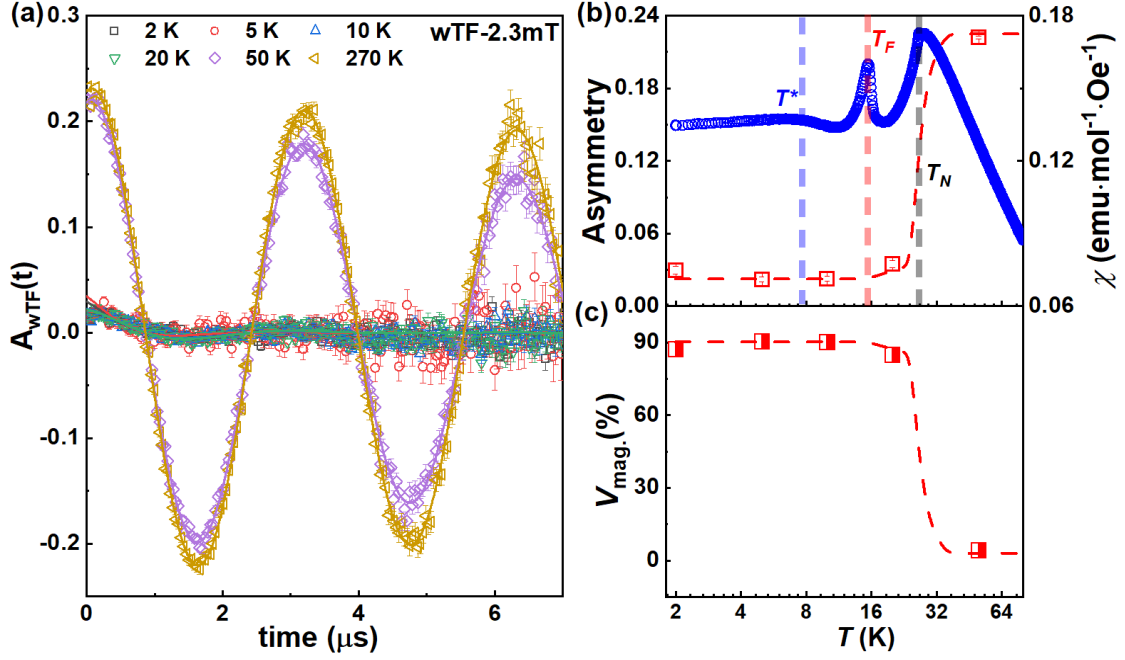

Figure 3. **Magnetic phase transition and magnetic volume fraction of NCTO, as measured by wTF-μSR.** (a) Time-domain wTF-μSR spectra collected at different temperatures in a weak transverse field of 2.3 mT, acquired at the M20D spectrometer. Solid lines are the fit results utilizing Eq. (1). To highlight the low-amplitude oscillations below $T_N$, the respective time-domain wTF-μSR spectra are also plotted in Fig. S7 [42]. (b) The temperature-dependent $A_{NM}$ asymmetry (left axis and red symbols) was obtained from fits of the wTF-μSR data. For a comparison, we present the zero-field cooling (ZFC) magnetic susceptibility curve $\chi(T)$ at $B = 0.01$ T (right axis and blue symbols) with $\boldsymbol{B}$ // $\boldsymbol{a^*}$-axis. The magnetic susceptibility data were taken from Ref. [19]. (c) The estimated magnetic volume fraction versus temperature. The red dashed lines in (c) and (d) are the guides for the eyes.

The analysis of wTF-μSR spectra allowed us to establish the temperature evolution of the magnetic volume fraction and to determine the magnetic transition temperatures. Here, an external field of $B = 2.3$ mT, applied perpendicular to the initial muon-spin direction, leads to a precession of the muon spins with a frequency $\gamma_\mu B$ (where $\gamma_\mu = 2\pi \times 135.53$ MHz/T is the muon's gyromagnetic ratio), as shown in Fig. 3(a). Note that a field of 2.3 mT is much smaller than the internal fields created by the long-range magnetic order in the NCTO single crystal (see Fig. 4). Therefore, the muon-spin precession reflects only the non-magnetic part of the sample, since in its long-range magnetic order the sample exhibits a very fast muon-spin depolarization, here, in less than a few tenths of μs. Consequently, without considering this very fast relaxation, the wTF-μSR spectra can be described by the function:

$$A_{wTF}(t) = A_{NM} \cos(\gamma_\mu B_{int} \cdot t + \varphi) e^{-\lambda t}, \qquad (1)$$

where $A_{NM}$ is the initial muon-spin asymmetry reflecting the muons implanted in the nonmagnetic (NM) or paramagnetic (PM) fraction of NCTO single crystals; $B_{int}$ is the local field sensed by muons (here almost identical to the applied magnetic field); $\varphi$ is an initial phase and $\lambda$ is the muon-spin relaxation rate.

Figure 3(b) presents the temperature dependence of the asymmetry for the wTF-µSR spectra. Below $T_N$, a static spin component leads to a fast reduction of asymmetry. Hence, $A_{NM}$ starts to decrease quickly as one approaches the AFM ordering temperature, consistently with the magnetic susceptibility data. Another broad and weak peak was observed in the wTF-5mT data at temperatures close to $T_F$ (see Fig. S3) [42]. Although NCTO undergoes three successive AFM transitions [indicated by dashed lines in Fig. 3(b)], the $A_{NM}(T)$ curve does not capture the possible phase transition at $T^*$ [13]. This indicates that the internal field or the magnetic structure is not significantly modified close to $T^*$. In the case of a fully magnetically ordered material, the magnetic volume fraction $V_{mag}$ at zero temperature is close to 100% [57,58]. Here, the temperature dependence of the magnetic volume fraction was estimated from $V_{mag}(T) = 1 - A_{NM}(T)/A_{NM}(T>T_N)$ [see Fig. 3(c)]. In our case, the static magnetic volume fraction is up to 90% and a possible spin-glass behavior, typically revealed by ac susceptibility measurements, is absent in our NCTO single crystal [26]. These results indicate that, below $T_N$, NCTO can be considered as fully magnetically ordered, indicative of a high sample quality.

If the electronic magnetic moments fluctuate very fast (typically above $10^{12}$ Hz in the PM state), the muon-spin polarization would not be influenced. When the system starts to enter the magnetically ordered state, such fluctuations slow down significantly. This scenario gives rise to a fast depolarization with superimposed oscillations, reflected in the appearance of static magnetic moments and the coherent precession of the muon polarization, respectively. Figures 4(a) and (b) show the ZF-µSR time spectra for both $S_\mu$ // $a$ and $S_\mu$ // $c$-axis (see also Fig. S4 [42]). Such fast depolarization (within 50 ns) with superimposed oscillations below $T_N$, usually occurs in some strongly frustrated antiferromagnets [58,59], thus suggesting the frustrated magnetic ground state of NCTO originating from the Kitaev-type frustration [13,33-35,41]. To track the evolution of the muon asymmetry across the whole temperature range, the ZF-μSR spectra were modeled by:

$$\mathbf{A}_{ZF}(t) = A_1 \cdot \left[ \alpha \cos(\gamma_\mu B_{int} t + \varphi) \cdot e^{-\lambda_T t} + (1-\alpha) \cdot e^{-\lambda_L t} \right] + A_2 e^{-\lambda_{tail} t} \cdot G_{KT}. \qquad (2)$$

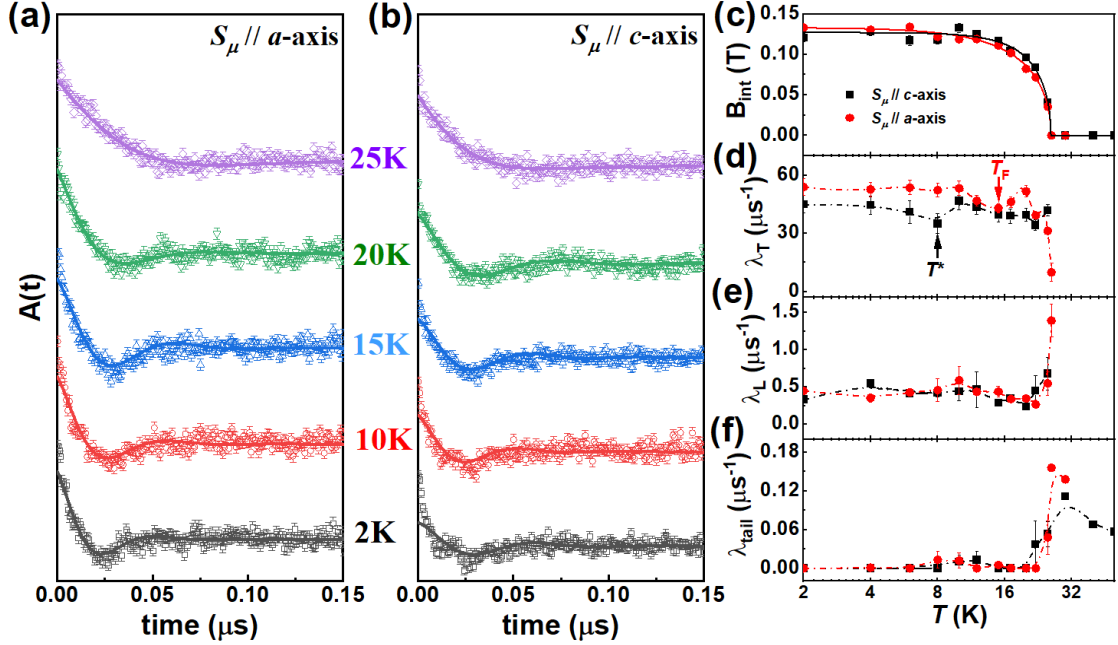

Figure 4. **Zero-field μSR spectra of NCTO.** (a) and (b) Zero-field (ZF) μSR time spectra measured at selected temperatures (displaced vertically for clarity) with the muon-spin direction $S_\mu$ // $a$ and $c$-axis, respectively. The solid curves represent fits to Eq. (2). (c)-(f) Temperature dependence of the $B_{int}$, $\lambda_T$, $\lambda_L$, and $\lambda_{tail}$, respectively, as derived from ZF-μSR analysis. Solid lines in panels (a) are fits using a phenomenological equation described in the text; dash-dotted lines in panels (d)-(f) are guides to the eyes.

Here, $\alpha$ and $1 - \alpha$ are the oscillating (i.e., transverse) and nonoscillating (i.e., longitudinal) fractions of the μSR signal, $\lambda_T$ and $\lambda_L$ represent the transverse and longitudinal muon-spin relaxation rates, $A_1$ and $A_2$ represent the asymmetries of the two nonequivalent muon-stopping sites. The muons stopping at the second site do not undergo any precession but show only a weak relaxation, which can be described by an exponential relaxation $\lambda_{tail}$. The ZF-μSR spectra show only a weak relaxation at temperatures above $T_N$. The $G_{KT}$ is the static Kubo-Toyabe relaxation function, normally used to describe the muon-spin relaxation due to the nuclear moments. In the magnetically ordered state, the muon-spin relaxation due to the electronic moments is significantly larger than the contribution from the nuclear moments. Therefore, below $T_N$, the $G_{KT}$ term can be safely ignored. In our analysis, we set $G_{KT}$ to 1 at temperatures below $T_N$ (see details in Fig. S5) [42].

Usually, changes in the magnitude of the magnetic moment can be detected by the temperature-dependent internal magnetic field $B_{int}(T)$. However, the modulation of magnetic domains is related to the distribution of internal magnetic fields. The fit parameters of the ZF-μSR spectra, here summarized in Fig. 4(c)-(f), help us to distinguish the origin of these complex magnetic orders. It is worth mentioning that, as the temperature changes, $\alpha$ was let to vary, so as to ensure a more reasonable parameter set below $T_N$. Indeed, the complex competition between the two different exchange frustrations of: Kitaev- and $J_1$-$J_3$-type [13,33-35,41], can trigger significant spin fluctuations accompanied by changes in spin directions, as shown in Fig. S6(c) [42]. The temperature-dependent parameter $\alpha$ shows weak anomalies near $T_F$ and $T^*$, most likely reflecting the modulation of the magnetic domains and the change of electronic moment directions are moderate.

The $B_{int}(T)$ curves for $\boldsymbol{S_\mu}$ // $\boldsymbol{a}$ and $\boldsymbol{S_\mu}$ // $\boldsymbol{c}$ exhibit similar features [see Fig. 4(c)], reflecting the local ordered magnetic moment of the $Co^{2+}$ ions and giving one distinct phase transition temperature $T_N$. Both curves are consistent with the temperature evolution of the magnetic moments obtained from powder [39] and single-crystal neutron diffraction (see details in Fig. 1). A phenomenological equation $B_{int}(T) = B_{int}(0K) \cdot \left[1 - (T/T_N)^\gamma\right]^\delta$ describes very well the $B_{int}(T)$ curves shown in Fig. 4(c). Here, $B_{int}(0K)$ is the internal magnetic field at 0 K, $\gamma$ and $\delta$ are two empirical parameters. The fitted parameters are summarized in Table SI in the supplementary materials [42].

Interestingly, the other two anomalies at $T_F$ and $T^*$, are clearly visible in the temperature-dependent $\lambda_T(T)$ curves, with both $\boldsymbol{S_\mu}$ // $\boldsymbol{a}$ and $\boldsymbol{S_\mu}$ // $\boldsymbol{c}$. We recall that the decay rate $\lambda_T$ reflects the width of the static magnetic field distribution at the muon-stopping site [see Fig. 4(d)]. These results suggest that the anomalies observed at $T_F$ and $T^*$ should be attributed to a redistribution of the magnetic domains originating from three different *k*-vectors of the zigzag AFM order. Hence, the multi-domain structure is compatible with both the ZF-μSR results and the single-crystal neutron diffraction data.

Spin fluctuations can be traced by the longitudinal relaxation rate $\lambda_L$. As shown in Fig. 4(e), both $\lambda_L(T)$ curves diverge near $T_N$, but also drop significantly below $T_N$, indicating that spin fluctuations are the strongest close to the onset of the AFM order. In NCTO, $\lambda_L$ becomes approximately constant (0.45 μs$^{-1}$) at low temperatures below $T_N$. In YbMgGaO$_4$, another promising QSL candidate material, a similar *T*-independent behavior, with a low-temperature value of 0.3 μs$^{-1}$, was suggested to reflect its very strong quantum fluctuations [60]. Hence, the temperature independence of $\lambda_L$ might indicate strong quantum fluctuations also in NCTO. Incidentally, below $T_N$, the temperature-dependent $\lambda_{tail}(T)$ is nearly zero [see Fig. 4(f)], here reflecting a simple exponential correction originating from the low background.

We now discuss the zigzag AFM ground state with a multi-domain structure, accompanied by strong quantum fluctuations in NCTO. By using the same fitted parameters, $T_N \approx 26$ K and $\beta \approx 0.22$, the power-law function $I(T) = I_0 \cdot [1-(T/T_N)^2]^\beta$ gives similar $I(T)$ curves at the $M$ and $M_1$ points [see Fig. 1(c)]. However, the small bump at $T_F$, corresponding to an opening of the spin-wave-excitation gap [31], is observed only in the $I(T)$ curve measured at the $M_1$-point. The absence of an anomaly at $T_F$ in the $I(T)$ curve of the $M$ point is consistent with the magnetic anisotropic-exchange interactions inducing non-equivalent magnetic domains at the $M$ and $M_1$ points. The field-dependent intensities of the magnetic reflections further confirm that the magnetic domains oriented along the magnetic-field direction grow, while those along other directions are suppressed. At the same time, the $\lambda_T(T)$ curves suggest that a domain reorientation occurs at $T_F$ and $T^*$. Our results indicate that the most relevant anisotropic interactions, $K$ and $\Gamma$ [13,33-35,41], are responsible for the highly anisotropic response in a variety of complex phases. They lead to the motion of the magnetic domains and support the multi-domain structure of NCTO.

Based on the generalized Heisenberg–Kitaev model [13,19,28,33-35,41], many previous studies have revealed a complex competition between the two types of exchange frustrations: Kitaev and $J_1$-$J_3$, which may lead to strong quantum fluctuations in the ground state of NCTO. Such quantum fluctuations are clearly evidenced by our ZF-µSR measurements: 1) NCTO exhibits a very fast depolarization (within 50 ns), typical of strongly frustrated antiferromagnets [58,59] [Fig. 4 (a) and (b)]; 2) The temperature independence of $\lambda_L$, here close to 0.45 µs$^{-1}$ [Fig. 4 (e)] at low temperatures, is similar to that observed in YbMgGaO$_4$, known to exhibit strong quantum fluctuations under zero field [60]. Taken together, our µSR data indicate that the strong spin dynamics at low temperatures is of quantum origin, with the opening of a spin-wave-excitation gap below $T_F$ further eliminating the contribution of thermal fluctuations [31]. The zero-field quantum fluctuations inevitably remind us of the recently reported field-induced Kitaev QSL between 7.5 T and 10 T in NCTO [13,19,26,37]. Further, the broadening at the K-point, where the FWHM gradually increases with the magnetic fields [see inset in Fig. 2(c)], suggests that quantum fluctuations are enhanced by the applied magnetic fields. These results indicate that the applied magnetic fields could quickly suppress the magnetically ordered states and highlight the contribution of Kitaev interactions to quantum fluctuations.

In summary, we investigated both the magnetic structure characteristics and the ground-state dynamics in NCTO. Our most significant finding is that NCTO hosts a multi-domain zigzag AFM order with strong quantum fluctuations. Our results provide experimental evidence in favor of the Heisenberg–Kitaev model, where the coexistence of static magnetic order (from the non-Kitaev interactions) with dynamic quantum fluctuations (from the frustrated $K$ term) suggest a highly frustrated magnetic structure.

*Note added*. While preparing the present manuscript, we noticed that µSR experiments, with a lower early-time resolution (here up to 50 ns), were presented independently [61].

## ACKNOWLEDGMENTS

G.T.L., T.S. and J.M. thank the financial support from the National Key Research and Development Program of China (No. 2022YFA1402702), the National Science

Foundation of China (Nos. U2032213, 12004243, 12374105). G.T.L. acknowledges support from the Startup Fund for Young Faculty at SJTU (SFYF at SJTU). J.M. thanks the interdisciplinary program Wuhan National High Magnetic Field Center (No. WHMFC 202122), Huazhong University of Science and Technology. T.S. acknowledges support from the Natural Science Foundation of Shanghai (Grant Nos. 21ZR1420500 and 21JC1402300), Natural Science Foundation of Chongqing (Grant No. CSTB-2022NSCQ-MSX1678). The work performed in the University of Tennessee (crystal growth) was supported by NSF-DMR-2003117. We acknowledge the neutron beam time from SINQ with Proposal 20222504, SNS with Proposal No. IPTS-27393.1, and ANSTO with Proposal No. P15604. We thank ACNS for the beam time and the sample environment group at ACNS for the support. This research used resources at the Spallation Neutron Source, a DOE Office of Science User Facility operated by the Oak Ridge National Laboratory. Part of this work is based on experiments performed at the Swiss spallation neutron source SINQ, and the Swiss muon source at the Paul Scherrer Institut, Villigen, Switzerland.

**Supplementary Material for "Static magnetic order with strong quantum fluctuations in spin-1/2 honeycomb magnet Na$_2$Co$_2$TeO$_6$"**

*Inelastic neutron scattering.* For the experiments on SEQUOIA time-of-flight spectrometer at the Spallation Neutron Source, Oak Ridge National Laboratory, USA, measurements at 2 K with applied field $B$ = 0 T and 4 T were performed by rotating the sample around the vertical axis (sample $a^*$-axis) in steps of 1° with $E_i$ = 18 meV and choppers in high-resolution mode, yielding a full-width at half-maximum (FWHM) elastic energy resolution of about 0.41 meV. In order to subtract the background, the INS data were collected at 90 K. The constant-$Q$ scans of the $K$-point (1/3, 1/3, 0) at 1.5 K with applied field $B$ up to 4 T and B // [-1,1,0] direction were measured on the Cold Triple Axis Spectrometer SIKA at the ANSTO, Australia. For the measurements on SIKA, data were collected using a fixed final-energy mode with $E_f$ = 5.0 meV.

*Muon spin rotation or relaxation.* For the experiments on M20D surface muon beam line at TRIUMF, the aligned NCTO crystals were positioned on a thin silver tape, Fig. S2(a) and (b), with their $c$ axes parallel to the muon momentum direction. We aimed at studying the temperature evolution of the magnetically ordered phase and the dynamics of spin fluctuations. For the experiments on GPS instrument, since NCTO is thin and flat, we simply used two layers of samples and then wrapped them with Kapton foil, Fig. S2(c) and (d). The muon momentum was always parallel to the $c$-axis of the crystal. The muon spin should be tuned in the $ab$-plane (TRAN mode) or along c-axis (LONG mode). Based on the wTF-μSR data, we could determine the temperature evolution of the magnetic volume fraction. For the wTF-μSR measurements, the applied magnetic field was perpendicular to the muon-spin direction.

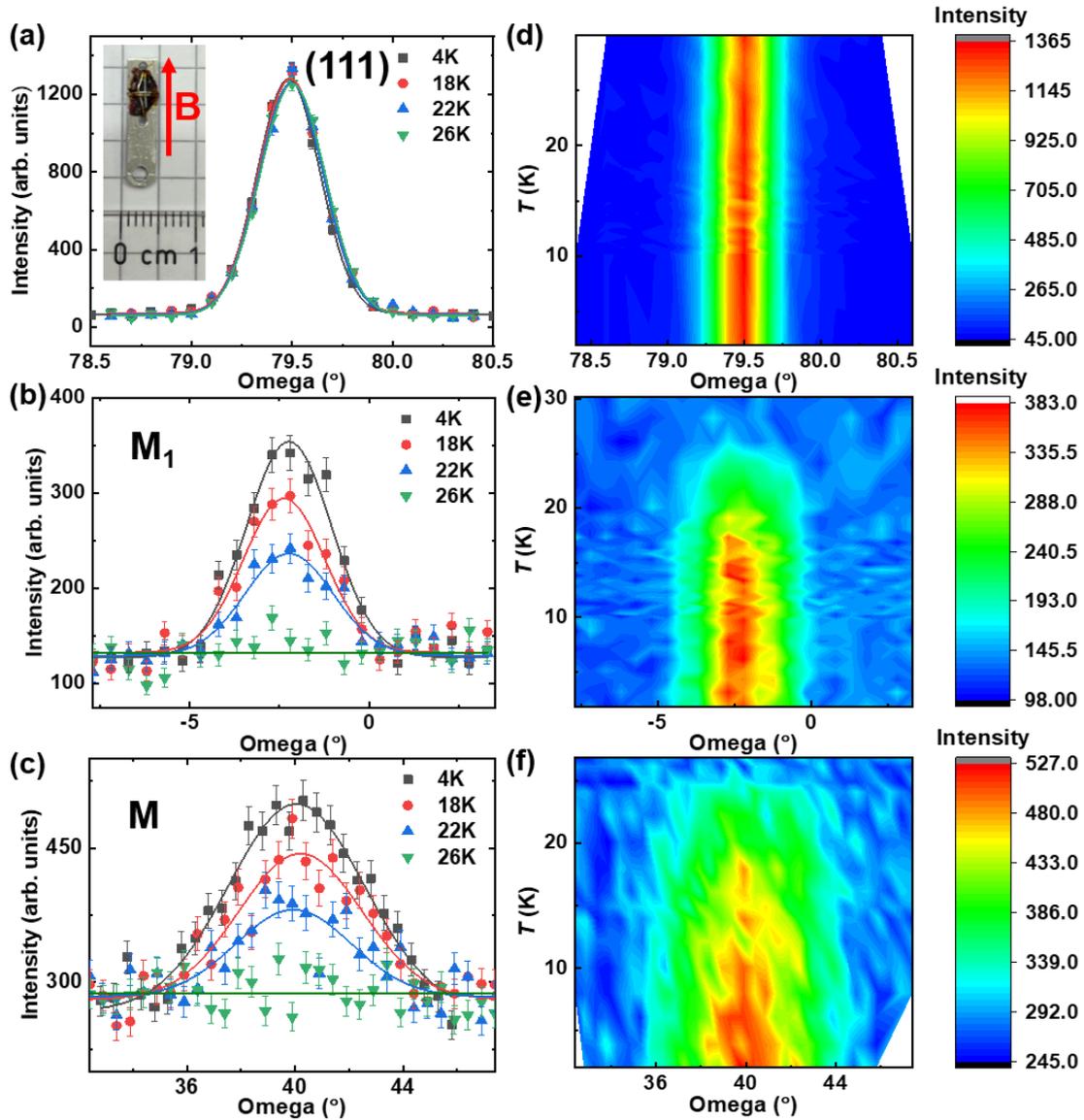

Figure S5. Single-crystal neutron diffraction of NCTO at the ZEBRA diffractometer. (a), (b), and (c) Zero-field temperature dependence of the nuclear peak (111), magnetic peak $M_1$ (-0.5, 0.5, 1) and M (0.5, 0, 1) at selected temperatures. The data collected at each temperature were fitted by a Gaussian function (the solid lines) to obtain an integrated intensity plotted in Figure 1. Inset: the sample used for the single crystal neutron diffraction experiment with the magnetic field applying along [-1, 1, 0] direction. (d), (e), and (f) The contour maps of (111), (-0.5, 0.5, 1), and (0.5, 0,1) peaks. Apparently, there is no lattice distortion or ferromagnetic contribution with temperature evolution from (d).

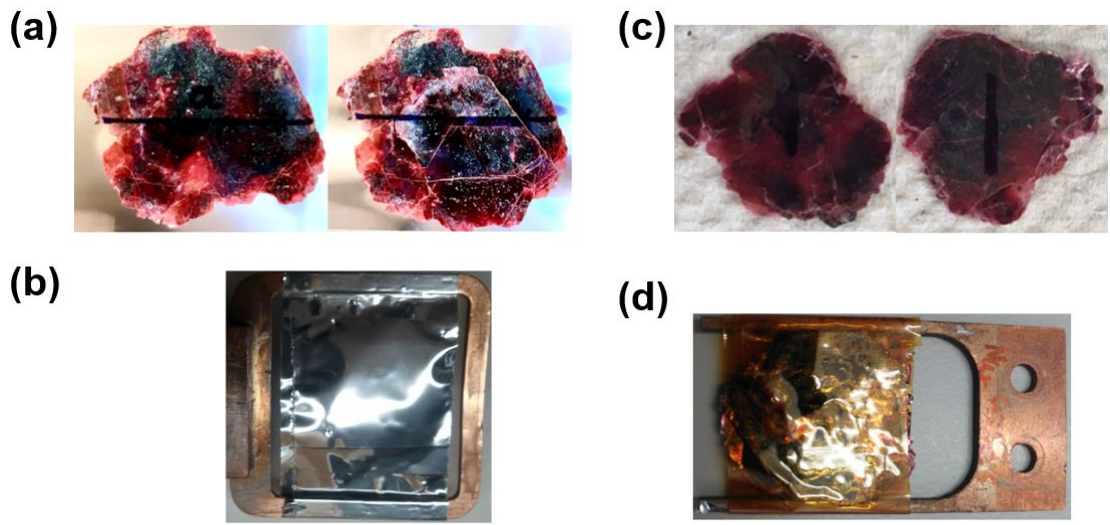

Figure S6. Photos of the samples measured via μSR. (a) Sample measured at the M20D spectrometer. The scribed lines on the samples mark the *a*-axis of the crystal. (b) The aligned NCTO crystals are positioned on a thin silver tape for M20D spectrometer. (c) Photos of the sample measured at the GPS spectrometer. (d) In this case, the aligned NCTO crystals were wrapped with a Kapton foil. The scribed lines on the samples mark again the *a*-axis of the crystal.

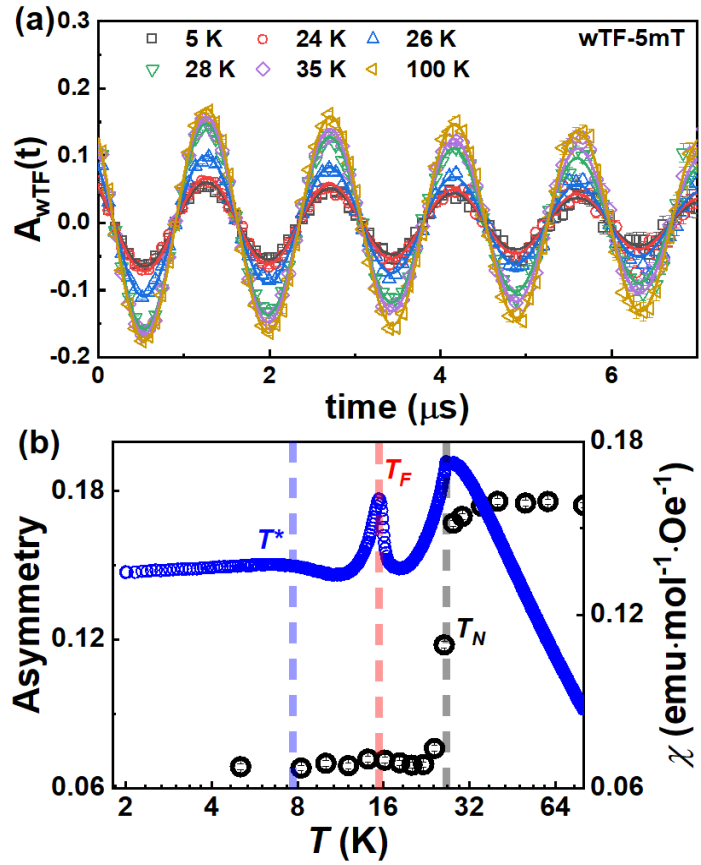

Figure S7. (a) Time-domain wTF-μSR spectra collected at different temperatures in a weak transverse field of 5 mT, as acquired at the GPS spectrometer. The solid lines represent the fit results by the Eq. (1). (b) Temperature-dependent $A_{NM}$ asymmetry (left axis, black and red curves) as obtained from fitting the wTF-μSR spectra. For comparison, we present the zero-field cooling (ZFC) magnetic susceptibility $\chi(T)$ curve at $B = 0.01$ T (right axis and blue curve) with $\boldsymbol{B}$ // $\boldsymbol{a^*}$-axis; The magnetic susceptibility data were taken from Ref. [19] of the main text.

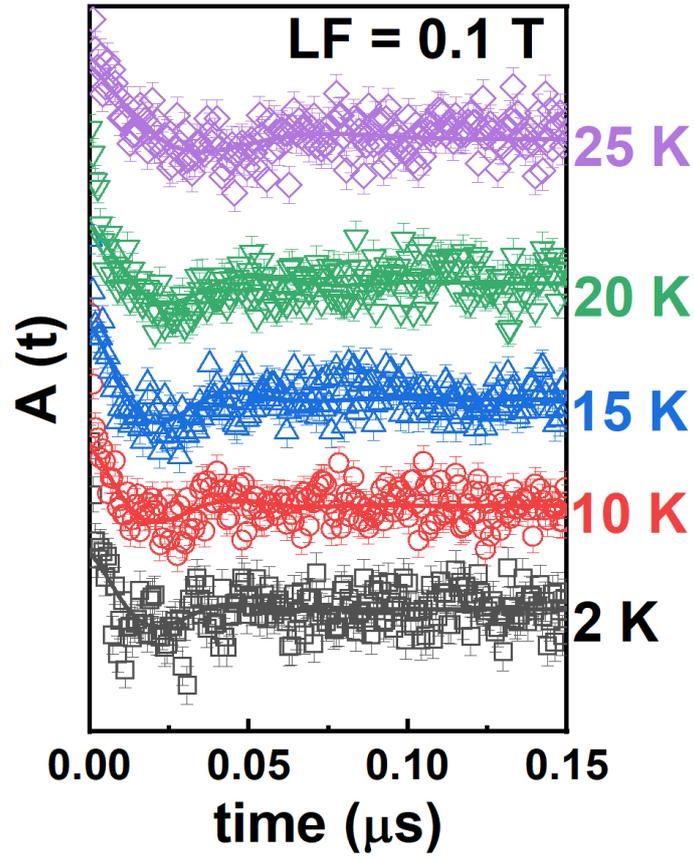

Figure S8. The μSR time spectra with LF = 0.1 T measured at selected temperatures (spectra displaced vertically for clarity) with the muon-spin parallel to the *c*-axis. Solid curves represent the results of least-squares fits using Eq. (2) in the main text. The fast depolarization with superimposed oscillations within 50 ns (originating from the AFM order) are confirmed by the μSR time spectra with LF = 0.1 T.

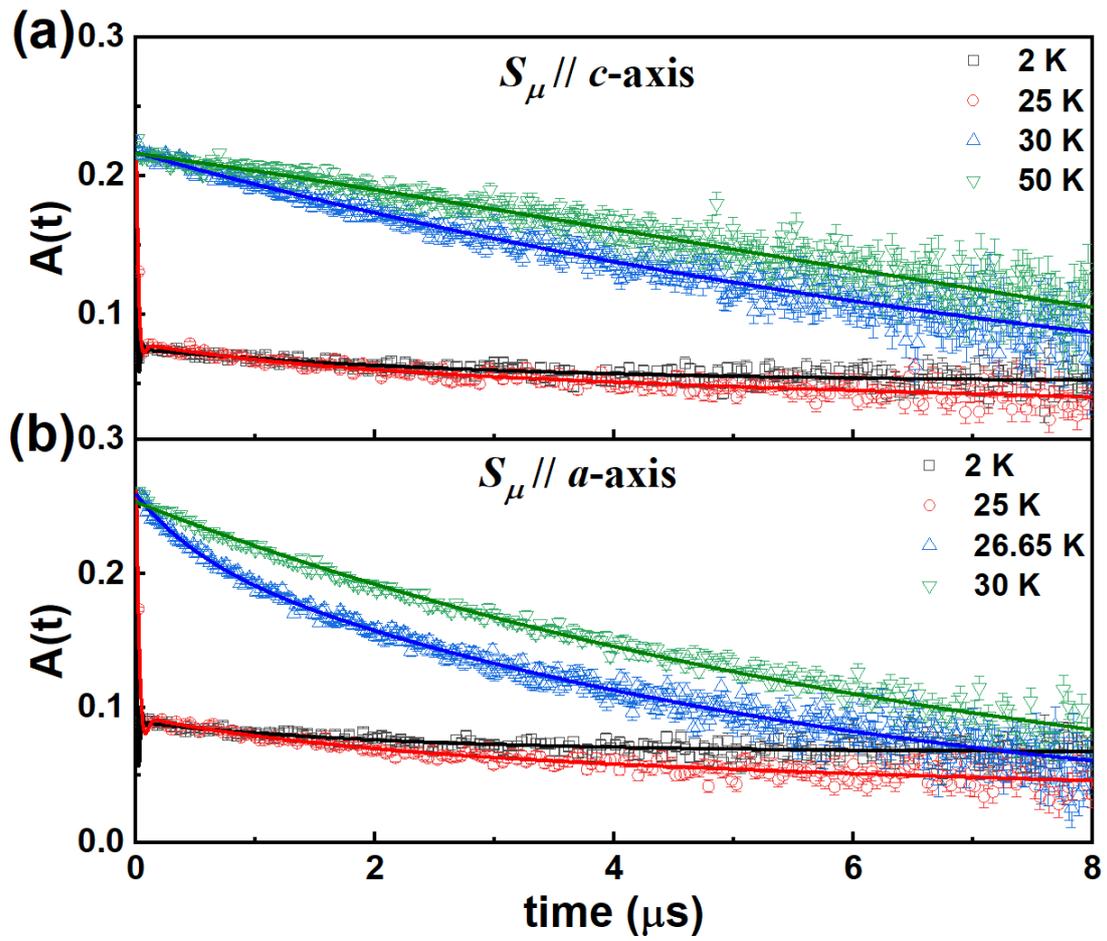

Figure S9. Zero-field μSR data of $Na_2Co_2TeO_6$ at selected temperatures, from the high-temperature paramagnetic state to the antiferromagnetic ground state. Data are well described by Eq. (2) in the main text.

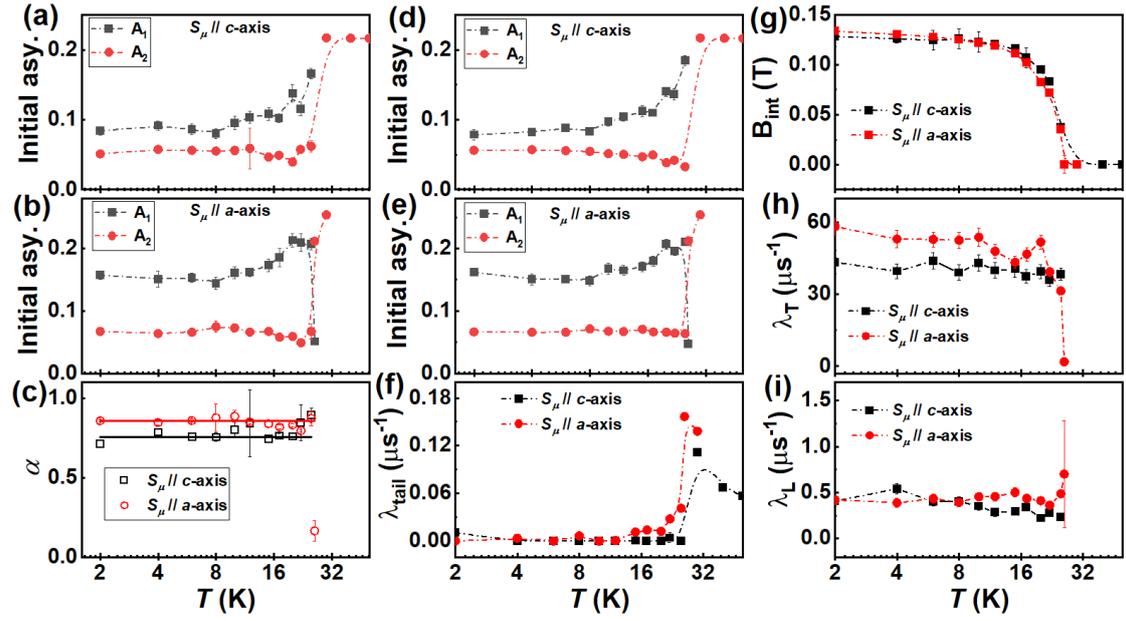

Figure S10. The fitted parameters by Eq. (2) in the main text. (a) and (b) The temperature-dependent initial asymmetry $A_1$ and $A_2$ are obtained by using unconstrained α values in Eq. (2), as shown in (c). (d)-(i) Temperature dependence of the initial asymmetry, $\lambda_{\text{tail}}$, $B_{\text{int}}$, $\lambda_T$, and $\lambda_L$, respectively, as derived from ZF-μSR analysis with constant values of 0.755 for $\mathbf{S_\mu}$ // **c**-axis and 0.8567 for $\mathbf{S_\mu}$ // **a**-axis. The dash-dotted lines are guides to the eyes.

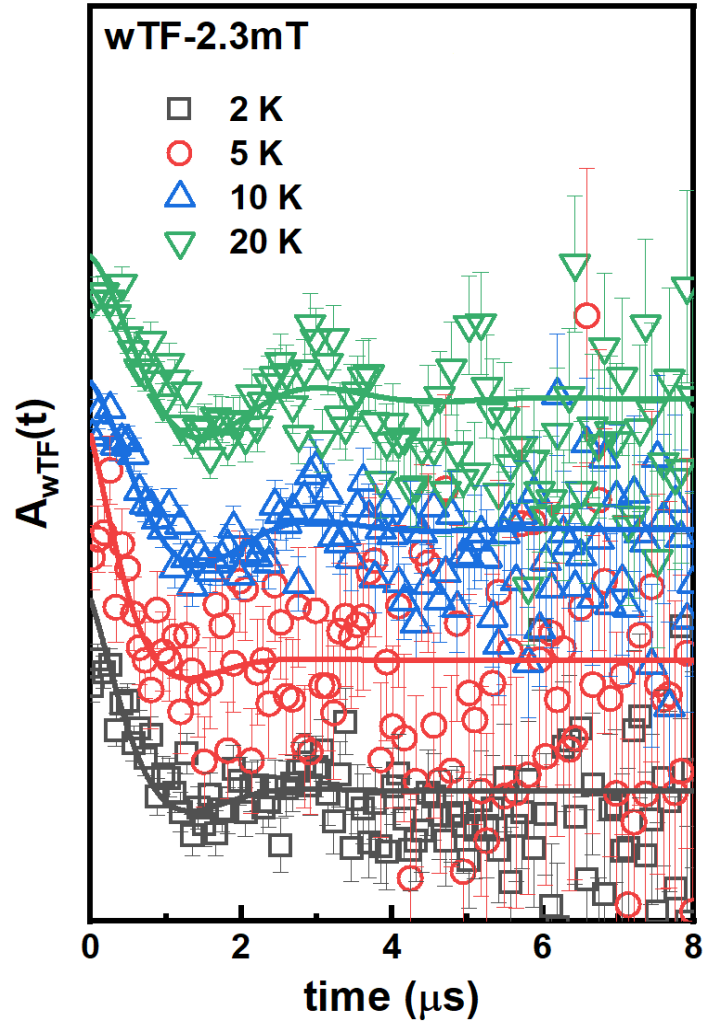

Figure S11. Time-domain wTF-μSR spectra collected at M20D in a weak transverse field of 2.3 mT below $T_N \sim 26$ K. Solid lines represent the fit results by means of Eq. (1) in the main text. Due to the weak oscillation, it is difficult to maintain consistency. Hence, we fixed the phase and the internal field below $T_N$ to those of the 50-K dataset, for which the phase was -9.81 degrees and the internal field 2.362 mT.

TABLE SI. Summary of the NCTO single-crystal parameters as obtained from magnetization and µSR measurements.

| | $\chi(T)$ | ZF-µSR $S_\mu$ // a-axis | ZF-µSR $S_\mu$ // c-axis |
|---|---|---|---|
| $T_N$ (K) | 26 | 25.8 | 25.8 |
| $T_F$ (K) | 15 | 15 | / |
| $T^*$ (K) | 7 | / | 8 |
| $B_{int}$ (T) | / | 0.13 | 0.127 |
| $\gamma$ | / | 1.93 | 3.16 |
| $\delta$ | / | 0.47 | 0.47 |